\documentstyle[epsf]{mn}

\def\Gyr{{\rm\,Gyr}}

\def\kms{{\rm\,km\,s^{-1}}}
\def\kpc{{\rm\,kpc}}

\def\msun{{\rm\,M_\odot}}

\def\yr{{\rm\,yr}}

\def\spose#1{\hbox to 0pt{#1\hss}}
\def\lta{\mathrel{\spose{\lower 3pt\hbox{$\mathchar"218$}}
     \raise 2.0pt\hbox{$\mathchar"13C$}}}
\def\gta{\mathrel{\spose{\lower 3pt\hbox{$\mathchar"218$}}
     \raise 2.0pt\hbox{$\mathchar"13E$}}}

\def\pdrv#1#2{{\partial #1 \over \partial #2}}

\def\tdrv2#1#2{{d^2 #1\over d{#2}^2}}

\def\half{{1\over 2}}
\def\bfr{{\bf r}}
\def\bfv{{\bf v}}

\def\fbh{f_{bh}}

\def\vrel{v_{rel}}

\def\gd{\Gamma_{d}}
\def\Fd{f_{d}}
\def\Nd{N_{d}}
\def\Md{M_{d}}
\def\Mt{M_{t}}
\def\Mz{M_{0}}
\def\Mbh{M_{bh}}

\def\be{\begin{equation}}
\def\ee{\end{equation}}

\begin{document}

\title{Constraints on the mass and abundance of black holes in the
	Galactic halo: the high mass limit}

\author[Chigurupati Murali, Phil Arras and Ira Wasserman]
{Chigurupati Murali$^1$, Phil Arras$^2$ and Ira Wasserman$^2$\\ 
$^1$Canadian Institute for Theoretical Astrophysics, McLennan Labs, 
University of Toronto,\\ 60~St.\ George St., Toronto M5S 3H8, Canada\\
$^2$Center for Radiophysics and Space Research, Cornell University}\maketitle

\medskip

\begin{abstract}
We establish constraints on the mass and abundance of black holes in
the Galactic halo by determining their impact on globular clusters
which are conventionally considered to be little evolved.  Using
detailed Monte Carlo simulations, and simple evolutionary models, we
argue that black holes with masses $\Mbh\gta (1-3)\times 10^6\msun$
can comprise no more than a fraction $\fbh\approx 0.17$ of the total
halo density at Galactocentric radius $R\approx 8$ kpc. This bound
arises from requiring stability of the cluster mass function. A more
restrictive bound may be derived if we demand that the probability of
destruction of any given, low mass ($M_c\approx(2.5-7.5)\times
10^4\msun$) globular cluster not exceed 50\%; this bound is $\fbh\lta
0.025-0.5$ at $R\approx 8$ kpc.  This constraint improves those based
on disk heating and dynamical friction arguments as well as current
lensing results.  At smaller radius, the constraint on $f_{bh}$
strengthens, while, at larger radius, an increased fraction of black
holes is allowed.
\end{abstract}

\begin{keywords}
globular clusters: general -- Galaxy: halo -- Galaxy: structure --
dark matter
\end{keywords}

\section{Introduction}
What is the form and structure of dark matter in galactic halos?  A
variety of both baryonic and non-baryonic candidates exist (see Carr
1994 for a review of baryonic dark matter candidates) but there are
relatively few constraints so the question remains.

One longstanding suggestion is that of Lacey \& Ostriker (1985) who
proposed that halo dark matter consists of massive black holes with
$M_{bh}\sim 2\times 10^6 M_{\odot}$.  In so doing, they cast a
solution to two problems: 1) what is the composition of the dark
matter; 2) and what is the mechanism which heats the Galactic disk?
Their calculation showed that a steady flux of $2\times 10^6
M_{\odot}$ black holes passing through the disk would heat the disk in
the manner required to explain the velocity dispersion-age relation
$\sigma_*\propto t_*^{1/2}$ for disk stars (Wielen 1977).  

Although subsequent observational and theoretical work suggests that
an explanation of disk heating does not require massive black holes
(Carlberg et al 1985; Stromgren 1987; Gomez et al 1990; Lacey 1991)--
indeed, analysis of the disk heating problem is ongoing
(e.g. Sellwood, Nelson \& Tremaine 1998)-- one can, in any case, view
the disk heating argument as a disk heating {\it constraint}. The
constraint can be developed by generalizing the Lacey \& Ostriker
model to $M_{bh}>2\times 10^6 M_{\odot}$ with a less-than-unity
fraction of halo mass in black holes $f_{bh}<1$ (e.g. Carr, Bond \&
Arnett 1984; Wasserman \& Salpeter 1994).  Then, since the energy
input to the disk $\Delta E\propto M_{bh}^2$ for a single black hole,
any combination $f_{bh}M_{bh}\sim 2\times 10^6 M_{\odot}$ produces the
same net heating of the disk.  Therefore $f_{bh}M_{bh}\gta 2\times
10^6 M_{\odot}$ overheats the disk and is definitely not allowed.  The
generalization $f_{bh}<1$ is desirable, given the variety of dark
matter candidates, the results of microlensing surveys (e.g. Alcock et
al. 1997) and the fact that not all dark matter need be baryonic given
the bound from primordial nucelosynthesis (Pagel 1997).

Are there other constraints on the mass and abundance of black holes
in the Galactic halo?  In some sense, halo black holes are suprisingly
difficult to detect, given that there is considerable observational
evidence for black holes of similar mass ($10^6 M_{\odot} \lta M_{bh}
\lta 10^9 M_{\odot}$) in the centers of galaxies (e.g. Kormendy \&
Richstone 1995).  Conversely, they have been surprisingly difficult to
constrain or rule out.  Lacey \& Ostriker themselves remarked that the
accretion luminosity of such objects may be too high to have escaped
detection; however, no definitive constraint has been recorded (Carr,
Bond \& Arnett 1984; Carr 1994).  Hut \& Rees (1992) argued that
dynamical friction would drag $\sim 100$ of these objects into the
Galactic center in a Hubble time, leading to coalescence and
production of a central object much larger than allowed by
observational constraints ($M_{bh}\sim 2\times 10^6 M_{\odot}$; Genzel
et al 1997).  However, Xu \& Ostriker (1994) tested this argument with
detailed N-body simulations and found that typically only one black
hole would remain in the Galactic center due to three-body encounters.

Constraints from gravitational lensing are comparatively weak at this
time.  For the values of $M_{bh}$ considered in this paper, the large
size of the Einstein ring gives event durations orders of magnitude
too long for the present Galactic microlensing surveys.  Lensing of
quasars (Canizares 1982; Kassiola, Kovner, \& Blandford 1991)
restricts $\Omega_{bh}=\langle \rho_{bh} \rangle /\rho_c$ to be less
than about $10\%$, which is greater than the estimated mass in dark
haloes--an upper bound for the scenario we are considering.  Several
observing plans have been proposed to detect massive black
holes. Turner and Umemura (1997) argue that the Sloan Digital Sky
Survery and the Hubble Deep Field can place constraints on
$\Omega_{bh}$ by looking for extremely large amplifications of O and G
stars at cosmological distances.  Also, Turner, Wardle, \& Schneider
(1990) propose that $M_{bh} \sim 10^6 M_{\odot}$ black holes are
detectable through arcsecond size lensing of objects in M31 and the
Galactic Center.

Wielen (1985,1988) first pointed out that globular clusters also
constrain the properties of massive black holes in the Galactic halo
because of their susceptibility to external heating and tidal
disruption.  Later, Moore (1993) applied the same arguments to a set
of low mass globular cluster in the halo.  These arguments have been
re-examined in more detail by Klessen \& Burkert (1995) and Arras \&
Wasserman (1998), who first included cluster evolution due to black
hole heating in examining the constraints.  Our goal in this and
subsequent work will be to re-examine constraints on $f_{bh}$ and
$M_{bh}$ imposed by globular clusters using Fokker-Planck calculations
of cluster evolution which include the effects of encounters with
massive black holes.

Both Wielen (1985,1988) and Arras \& Wasserman (1998) delineate two
mass regimes for black holes: the low $M_{bh}$ regime, in which
individual collisions perturb a cluster only weakly, but where many
such collisions produce a steady, diffusive energy input; and the high
$M_{bh}$ regime, where a single encounter can destroy the cluster.  In
the low-mass limit, one can obtain limits on the product
$f_{bh}M_{bh}$; in the high-mass limit, one can obtain limits only on
$f_{bh}$ for $M_{bh}>M_{high}$.  Applying these arguments to Moore's
(1993) cluster sample, Arras \& Wasserman (1998) concluded that
$f_{bh}M_{bh}\lta 10^3 M_{\odot}$ in the low-mass regime and
$f_{bh}\lta 0.3$ in the range $10^6 M_{\odot}\leq M_{bh}\leq 10^7
M_{\odot}$.

Although Arras \& Wasserman (1998) included evolution due to black
hole heating, they did not consider the influence of internal
relaxation and post-collapse evolution in their calculations.  Thus
they pointed out the need for improved calculations to derive the most
robust constraints on $f_{bh}$ and $M_{bh}$ using the globular cluster
argument.

To address this issue in the present work, we combine the statistical
framework developed by Arras \& Wasserman (1998) with the
Fokker-Planck evolutionary calculations employed by Murali \& Weinberg
(1997a-c).  Our calculations include two-body relaxation, post
core-collapse evolution and tidal shocking by massive black holes.
Using a Monte Carlo approach, we focus on the high-mass regime and
directly compute the probabilities for strong collisions between
globular clusters and massive black holes.

In order to translate these calculations of the evolution of {\it
individual} clusters into bounds on $\fbh$, we need to consider the
implications of our results for the evolution of a {\it population} of
clusters. To accomplish this, we adopt two different points of
view. The first, and more restrictive, viewpoint is that global
studies of the evolution of the globular cluster population are
consistent with observations assuming {\it no} black holes at all
(e.g. Murali \& Weinberg 1997b), so that including black holes should
have only a minimal effect. Following this approach, we find that to
ensure 50\% survival probability for globular clusters with masses
$M_c\approx(2.5-7.5)\times 10^4\msun$ at $R=8$ kpc, the fraction of
the halo in black holes with masses $\Mbh\gta (1-3)\times 10^6\msun$
must be $\fbh\lta 0.025-0.05$. This limit on $\fbh$ is between one and
two orders of magnitude stronger than the disk heating constraint at
this $\Mbh$, and would imply that black holes with $\Mbh\sim
10^6\msun$ are not a candidate for baryonic dark matter in galactic
halos.

Our second approach, which turns out to be systematically less
restrictive, examines the evolution and stability of the globular
cluster population in the context of a simple model. As we shall see,
the two principal effects of perturbations by black holes are to
destroy clusters outright, and to cause surviving clusters to lose
mass.  We model this simply via a partial differential equation that
includes mass loss via an advection term, as well as destruction. In
order to obtain results valid for black hole masses $\sim 10^6\msun$,
we must extend the tidal approximation employed throughout most of
this paper, so that non-destructive encounters at impact parameters
inside the tidal radius of a cluster may be taken into account. (The
extension of our calculations to include such penetrating encounters,
as well as more detailed evolutionary models, will be treated by us
elsewhere.) In the context of these models, we find that requiring the
observed mass distribution of Galactic globular clusters to be stable
over $\approx 10$ Gyr requires $\fbh\lta 0.17$ at $R=8$ kpc for
$\Mbh\gta 2\times 10^6\msun$. This bound, although not as tight as our
more restrictive (and more qualitative) limit on $\fbh$, still implies
that massive black holes cannot be the primary constituent of the halo
dark matter.

The plan of the paper is as follows. We summarize the framework for
determining constraints in \S 2. In \S 3 we determine which globular
clusters may be considered relatively unevolved over the age of the
Galaxy in the absence of bombardment by black holes, and then find the
probability that they are destroyed for given values of $M_{bh}$ and
$f_{bh}$.  The results are extended to different cluster and black
hole parameters using scaling arguments.  We then discuss the
properties of clusters which are not destroyed outright, but instead
undergo many non-destructive collisions over the age of the galaxy.
Lastly, in \S 4 we discuss how our results constrain $M_{bh}$ and
$f_{bh}$.

\section{Framework}
\label{sec:framework}
To re-examine constraints on $f_{bh}$ and $M_{bh}$ set by globular
clusters, we incorporate collisions and encounters with black holes
into multi-mass Fokker-Planck calculations of cluster evolution, which
include two-body relaxation and phenomenological binary heating of the
core.  Our code descends from that of Chernoff \& Weinberg (1990).  In
practice, we take clusters on circular orbits at $R=16\kpc$: this
minimizes the effect of relaxation and allows us to neglect Galactic
tidal heating while subjecting clusters to the approximate black
hole-flux crossing the disk.  In addition, this radius corresponds to
the spatial region containing many of the clusters in the sample used
by Moore (1993).  See Table 1 below for a list of input parameters for
the calculation.

Given that clusters evolve and some will vanish through evolution
(e.g. most recently Murali \& Weinberg 1997a-c; Gnedin \& Ostriker
1997; Vesperini 1997), it is important at the outset to establish
which clusters to use in setting constraints.  The lifetime for given
cluster orbit scales roughly with internal dynamical time $t_{dyn}$
and cluster mass $M_c$ as $t_{life}\propto t_{dyn}M_c$ for quasistatic
evolution.  Since $t_{dyn}$ scales with Galactocentric position due to
tidal limitation, low-mass clusters in the inner Galaxy are the first
to vanish.  In general, for any particular orbit, there is a minimum
mass cluster which survives to the present-day.  Clusters with
evaporation timescales less than a Hubble time $t_H$ do not provide
straightforward constraints on $f_{bh}$ and $M_{bh}$: clusters
currently at or below the minimum mass might have had considerably
larger {\it initial} mass.

It is important to note that the predictions from evolutionary
calculations appear quantitatively consistent with observations.  The
main uncertainty in the Fokker-Planck calculations is the core heating
term: calculations predict that clusters have high central densities
in the post core-collapse phase, while it is unclear precisely how
this relates to observations (e.g. Drukier, Fahlman \& Richer 1992).
Nevertheless, the predicted death rates are not wildly inconsistent
with observations and differences of opinion arise mainly over the
importance of evolution in clusters at the peak of the luminosity
function, $M_c\sim 10^5 M_{\odot}$ (Murali \& Weinberg 1997b; Gnedin
1997; Harris et al 1998; Kundu et al 1998).

With this in mind, we adopt a two-step approach to investigating
constraints on black hole masses: 1) we first determine cluster
initial conditions which do not strongly evolve in smooth halos in a
Hubble time; 2) we then immerse these clusters in halos with black
holes to investigate the possible constraints.  While cluster survival
is most likely for large $M_c$, significant perturbation by black
holes is most likely for small $M_c$.  We consider cluster masses near
the lowest $M_c$ that can survive for $10 \Gyr$ when $f_{bh}=0$.

\subsection{Dynamics of encounters}
\label{sec:dynamics}
We specify isotropic distributions of perturbing black holes using the
$f_{bh}$-$M_{bh}$ parameterization: therefore the local number density
of black holes $n_{bh}(R)=f_{bh}\rho_{halo}(R)/M_{bh}$.  This implies
the relative velocity distribution for encounters given in equation
(16) of Arras \& Wasserman (1998).

Encounters between clusters and black holes are predominantly
impulsive given the Galactic rotation velocity $V_c\sim 220\kms$.  For
a cluster at $R\sim 16\kpc$, the characteristic internal velocity
$v_{int}\sim 5\kms$.  For a cluster on a circular orbit at $V_c$ and a
random perturber drawn from a non-rotating, isothermal halo, the
typical relative encounter velocity $V_{rel}\sim V_c$.  For the black
hole masses and abundances considered below, the influence of impacts
at $\sim 10$ tidal radii, $r_t$, is small.  Even at this distance, the
timescales $r_t/v_{int}>> 10r_t/V_c$ so the probability of a
non-impulsive encounter is very small.

To perform the simulations described below, we incorporate individual
collisions between black holes and globular clusters into
Fokker-Planck calculations using the impulse approximation.  To
determine the effect of each encounter, we calculate the second-order
change in the distribution function (e.g. Murali \& Weinberg 1997a).
The method is similar to procedures used by Murali \& Weinberg (1997a)
and Gnedin \& Ostriker (1997) in studies of cluster evolution in the
Galactic tidal field.  However, in the appendix, we show that there is
a small error in the previous treatments.  Our current treatment
remedies this.

This approach represents a linearization of the full collision
problem.  In complete generality, the collision problem requires
simultaneous solution of the coupled, collisionless Boltzmann-Poisson
equations.  Linearization imposes a limited range of validity.  In the
appendix, we show that this approach is valid for $dM/M\lta 0.15$ in a
single encounter.  We terminate any calculation where $dM/M\geq 0.15$.

We adopt the Fokker-Planck approach, rather than, say, N-body
simulations because we can study evolution due to both internal and
external effects and because we need a computationally feasible method
to conduct the Monte Carlo simulations described below.  While N-body
simulations permit fully non-linear calculations of strong collisions,
it is difficult to include two-body relaxation, core heating which
leads to post core-collapse evolution and the effect of weak
encounters in which only small mass loss occurs.  N-body simulations
are also much too expensive to use in Monte Carlo simulations.

\subsection{High-mass limit}
\label{sec:high_mass}
Our analysis focuses on the {\it high-mass limit} for halo black holes
(e.g. Bahcall, Hut \& Tremaine 1980; Wielen 1985; Arras \& Wasserman
1998).  The limiting mass is defined as the mass for which a single,
tidal encounter at the typical relative velocity can destroy a
cluster.  For completeness, we sketch the definition of the high-mass
limit following the detailed discussion given by Arras \& Wasserman
(1998).

Let us assume that cluster destruction occurs when a strong collision
unbinds a fraction $dM$ of the total mass.  As shown in Appendix A,
the fractional mass loss in the impulsive tidal limit
\begin{equation}
\vert dM/M\vert \equiv f=K M_{bh}^2/V_{rel}^2b^4
\end{equation}
where $b$ and $V_{rel}$ are the impact parameter and relative velocity
of the collision.  The constant
\begin{equation}
\label{eq:Kdef}
K={\kappa G^2 \langle r^2\rangle\over \sigma^2}.
\end{equation}
The quantity $\langle r^2\rangle$ is the mean square radius of the
cluster, $\sigma$ is its one-dimensional central velocity dispersion
and $\kappa$ is a dimensionless constant depending only on its
structure; see Arras \& Wasserman 1998, equations (36) and (37).  Note
that, from the virial theorem, $\sigma^2\propto r_t^{-1}$ and that, by
tidal limitation in an isothermal sphere, $r_t\propto R^{2/3}$, so
that $K\propto R^2$, where $R$ is the Galactocentric radius.  

Rewriting this, we may define the `destructive radius':
\begin{equation}
b_d=\biggl({K M_{bh}^2\over V_{rel}^2 f_d} \biggr)^{1/4},
\end{equation}
where $f_d$ is the fractional mass loss leading to destruction.  Given
a black hole of mass $M_{bh}$ moving at velocity $V_{rel}$ with
respect to the cluster, an encounter with any impact parameter $b\leq
b_d$ leads to fractional mass loss $f\geq f_d$ from the cluster.  Of
course, since the black hole can travel with a range of relative
velocities in relation to the cluster, we may invert this relation to
define the range in $V_{rel}$ which leads to destructive encounters
within $r_t$:
\begin{equation}
\label{eq:vrelmax}
V_{rel}\leq \biggl({K M_{bh}^2\over f_d r_t^4}\biggr)^{1/2}\equiv V_{rel,max}
\end{equation}
In general, cluster properties depend on time, so that $b_d$ and
$V_{rel,max}$ will too.

We define the limiting mass $M_{bh}\equiv M_{high}$ to be the black
hole mass for which $f_d=0.15$ in an encounter at $b=r_t$ with
$V_{rel}= 2.5\langle V_{rel}\rangle$, where $\langle
V_{rel}\rangle\approx 1.47V_c$.  The factor of 2.5 is introduced to
ensure that the probability of non-destructive collisions inside $r_t$
is very small. (The factor of 2.5 is probably larger than it needs to
be: we estimate a probability of about $10^{-5}$ for a nondestructive
collision inside $r_t$ when $M_{bh}=M_{high}$ with this choice.)  Note
that $(b_d/r_t)^2=M_{bh}/M_{high}$ for this value of $V_{rel}$, and,
more generally, $(b_d/r_t)^2=2.5\langle V_{rel}\rangle M_{bh}
/V_{rel}M_{high})$.

\subsection{Monte Carlo enumeration of collision probabilities}
\label{sec:coll_prob}
To proceed, we introduce a framework for calculating the probability
that a cluster experiences an encounter with fractional mass loss
$f_d$ in a Hubble time.  For the black hole background specified
above, the rate of encounters with $f\geq f_d$ is
\begin{equation}
\pdrv{N_d}{t}=2\pi\int_0^{\infty}dV_{rel}\int_0^{b_d} db b
V_{rel} n_{bh}(R) F(V_{rel},t)\equiv\Gamma_d,
\end{equation}
where $F(V_{rel},t)$ is the relative velocity distribution of the
black hole population with respect to the instantaneous motion of the
cluster.  For a non-evolving cluster on a circular orbit, Arras \&
Wasserman (1998) show that
\begin{equation}
N_d=\pi T\rho_{bh}(R)\biggl({K\over f_d}\biggr)^{1/2}.
\end{equation}
Implicit in equation (6) are a number of noteworthy scalings: $N_d$ is
actually independent of $M_{bh}\geq M_{high}$ and $M_c$, and
$N_d\propto R^{-1}$ given $f_{bh}$.  In general, the probability that
a cluster suffers an encounter with $f\geq f_d$ in a Hubble time is
\begin{equation}
\label{eq:Pd_an}
P_d=1-\exp(-N_d).
\end{equation}
Qualitatively, $N_d\lta 1$ is required for a cluster to avoid a single,
destructive encounter with a black hole.

In the present work, our goal is to determine $P_d$ using realistic
calculations of cluster evolution: namely the Fokker-Planck solutions
discussed above.  As mentioned above, the method is linear and valid
only for $f\leq 0.15$.  Therefore, if we take $f_d=0.15$, then it is
most appropriate to call $P_d$ the {\it probability for a strong
collision}.  However, we also refer to $P_d$ as the disruption
probability according to its original definition.

Since we are concerned with the probability that a cluster does or
does not suffer {\it one} such collision, the range of possible final
cluster states is broad: in other words, the evolution is stochastic.
We therefore calculate collision probabilities using Monte Carlo
simulations.

Each simulation has 60 realizations.  The collision history for each
realization is obtained by direct sampling of the relative velocity
distribution and space density of black holes of mass $M_{bh}$ within
10 initial tidal radii of the cluster.  The disruption probability is
calculated directly from the fraction of the 60 runs in which an encounter
with $f\geq 0.15$ occurs.  Initially, each cluster has mass $M_c$,
King parameter $W_0$, galactocentric radius $R$ and additional
parameters (internal mass spectrum: Murali \& Weinberg 1997b,c) given
in Table \ref{tab:cl_params}.

\begin{table*}
\caption{Cluster Initial Conditions}
\label{tab:cl_params}
\begin{tabular}{lllc} 
\\ \hline
\multispan3 Structure \hfill&Adopted values\hfill\\ \hline
$M_c$&& total mass \hfill&$2.5\times 10^4, 7.5\times 10^4 M_{\odot}$ \\
$W_0$&& King concentration parameter \hfill&$W_0=3, 5, 7$\hfill \\
$R_c$&& cluster limiting radius \hfill&set to tidal limit $r_t$\hfill \\ 
$t_{rh}$&& initial half-mass relaxation time\hfill& $2-6\times 10^9\yr$\\ 
	\hline
\multispan3 Internal mass spectrum\hfill&\\ \hline
$\beta$&& mass spectral index: $N(m)\propto m^{-\beta}$ \hfill&
        $\beta=2.35$ (Salpeter)\hfill\\
$m_l$&& lower mass limit \hfill&$m_l=0.1 M_{\odot}$\hfill\\
$m_u$&& upper mass limit \hfill&$m_u=2.0 M_{\odot}$\hfill\\ \hline
\multispan3 Orbit: \hfill &circular at 16 kpc\hfill \\\hline
&&\hfill&\\
\end{tabular}
\end{table*}

\section{Constraints from globular clusters}
\label{sec:glob_constraints}

\subsection{Cluster evolution in a smooth halo:$f_{bh}=0$}
\label{sec:smooth}

For $f_{bh}=0$, our calculations include only relaxation and
post-collapse heating of the core.  Table \ref{tab:t_evap1} gives the
evaporation times $t_{evap}$ for clusters on circular orbits at
$R_{gc}=16\kpc$.  The evaporation time is roughly $10 \Gyr\equiv t_H$
for $M_c\sim 10^4 M_{\odot}$, independent of concentration and roughly
proportional to the initial mass of the cluster.  The evaporation
times scale as $t_{evap}(R)=t_{evap}(R/16\kpc)$ for an isothermal
halo.  The lack of dependence on concentration is also seen in
calculations presented by Lee et al (1991) and Gnedin \& Ostriker
(1997).

>From these results, we conclude that clusters with initial masses
$M_c=2.5\times 10^4 M_{\odot}$ and $7.5\times 10^4 M_{\odot}$ provide
appropriate specimens to study under bombardment by black holes: at
this radius, they have $t_{evap}>>t_H$ for $f_{bh}=0$; the low-mass
cluster lets us probe the lowest values of $M_{high}$ while the
high-mass cluster is farther from the evaporation boundary, thus
giving greater confidence in the conclusions.

\begin{table*}
\caption{Evaporation times at 16 kpc (in Gyr)}
\label{tab:t_evap1}
\begin{tabular}{lccc}
$M_c (M_{\odot})/W_0$&$3$&$5$&$7$\\\hline
$1\times10^3$&1.2&1.2&1.2\\
$2.5$&2.6&2.7&2.7\\
$5.0$&4.6&4.7&4.7\\
$7.5$&6.6&6.7&6.8\\
$1\times 10^4$&8.5&8.5&8.8\\
$2.5$&20&21&21\\
$5.0$&38&40&40\\
$7.5$&55&59&59\\
$1\times 10^5$&73&78&77\\\hline
\end{tabular}
\end{table*}

\subsection{Collision probabilities in lumpy halos: $f_{bh}>0$}
\label{sec:probs}
Our main calculations depict the evolution of King model clusters on
circular orbits at $R_{gc}=16\kpc$ with a range of concentration and
mass.  We take halos with a range of $f_{bh}$ and $M_{bh}=M_{high}$,
where $M_{high}$ depends on the initial concentration and mass of the
cluster.  Figures 1-3 compare the results of the Monte Carlo
calculations with the analytic predictions for the fixed cluster
potential (equation \ref{eq:Pd_an}).  There appear to be statistical
fluctuations in the Monte Carlo results as well as a small systematic
trend for 15\% collision probabilities to lie below the fixed cluster
predictions.  Error bars are 68\% confidence regions found using
Bayesian methods (as in Arras \& Wasserman 1998) and are only
statistical. Overall, however, the agreement is surprisingly good
given that the analytic prediction neglects changes in the cluster
potential due to internal evolution.

\begin{figure}
\label{fig:w3_pd_m2.5_r16}
\epsfxsize=20pc
\epsfbox[12 138 600 726]{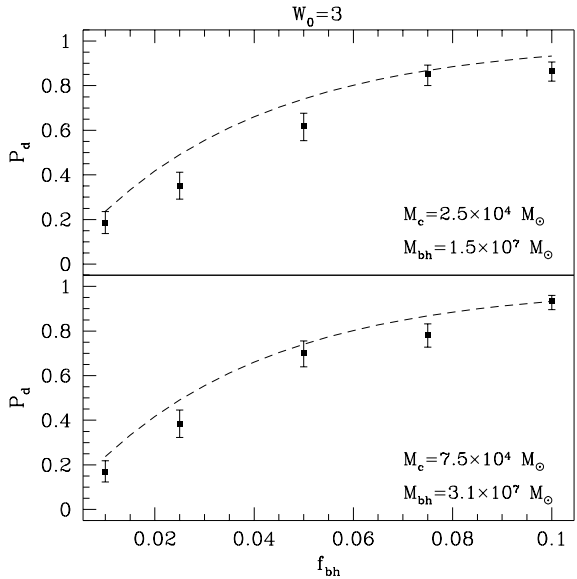}
\caption{Probabilities for 15\% encounters for $W_0=3$ clusters with
indicated masses at $16\kpc$ where $M_{bh}=M_{high}$.  Solid squares
with associated error bars show the results of Monte Carlo
simulations; dashed line shows the analytic prediction determined from
equations (6) and (7).}
\end{figure}

\begin{figure}
\label{fig:w5_pd_m2.5_r16}
\epsfxsize=20pc
\epsfbox[12 138 600 726]{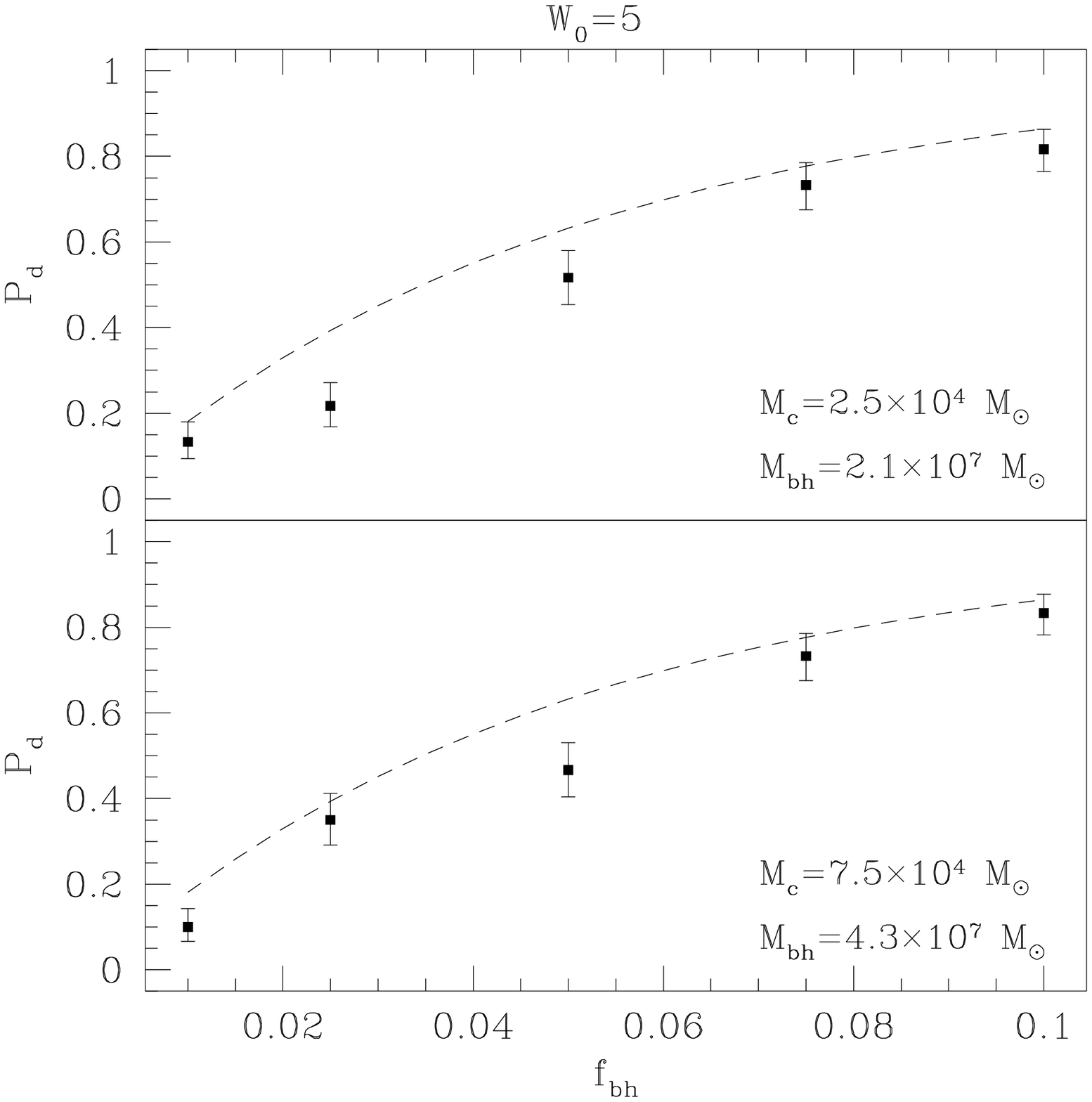}
\caption{Probabilities for 15\% encounters for $W_0=5$ clusters with
indicated masses at $16\kpc$ where $M_{bh}=M_{high}$.  Solid squares
with associated error bars show the results of Monte Carlo
simulations; dashed line shows the analytic prediction determined from
equations (6) and (7).}
\end{figure}

\begin{figure}
\label{fig:w7_pd_m2.5_r16}
\epsfxsize=20pc
\epsfbox[12 138 600 726]{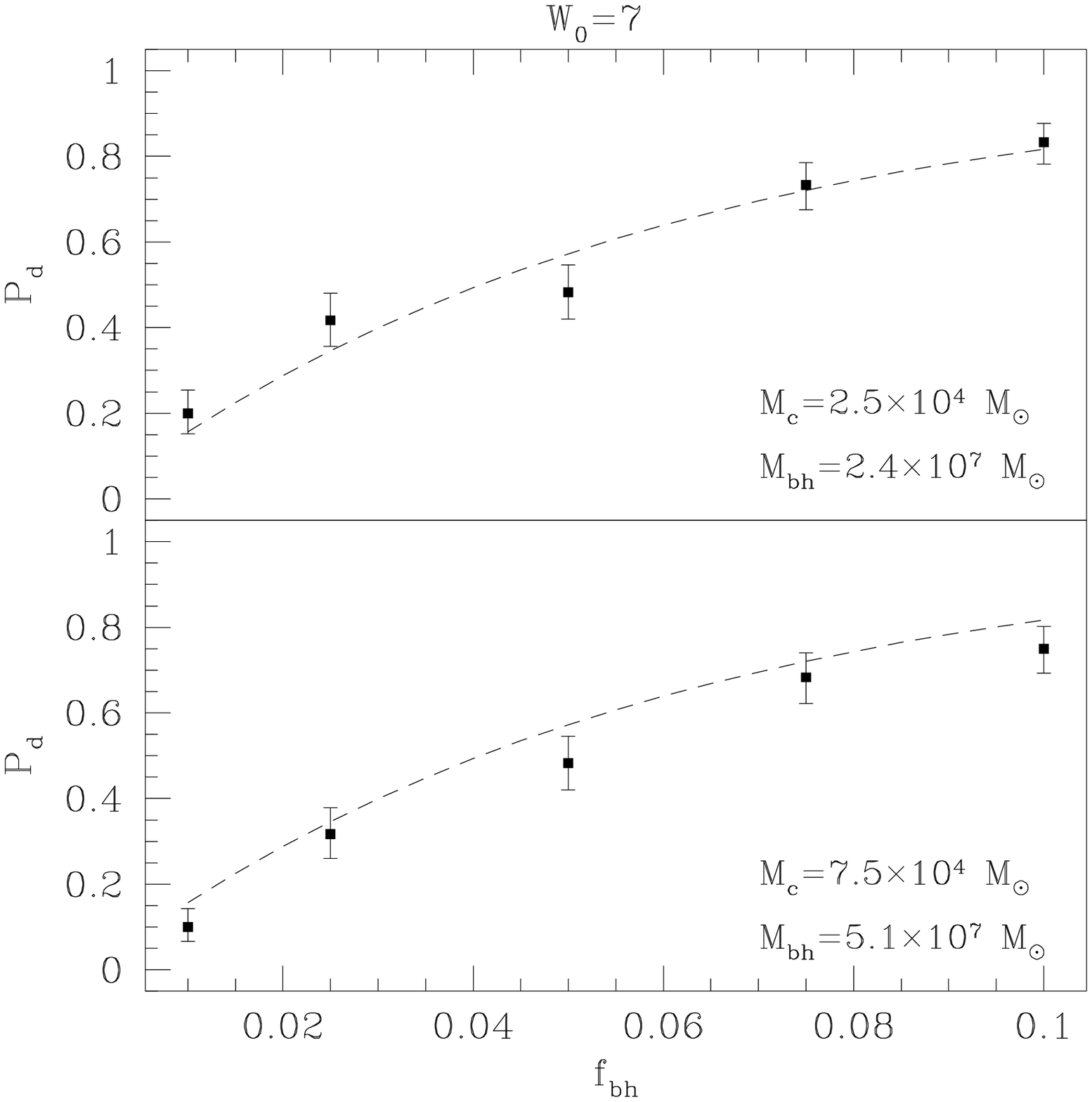}
\caption{Probabilities for 15\% encounters for $W_0=7$ clusters with
indicated masses at $16\kpc$ where $M_{bh}=M_{high}$.  Solid squares
with associated error bars show the results of Monte Carlo
simulations; dashed line shows the analytic prediction determined from
equations (6) and (7).}
\end{figure}

Lower $P_d$ might be expected in the simulations because two-body
relaxation hardens the potential while mass loss reduces the
cross-section for collisions.  Although evolution should help clusters
avoid strong collisions, $P_d$ is only reduced by roughly 10-20\% for
$f_{bh}\lta 0.1$.  Agreement is best at the smallest and largest
$f_{bh}$.  Agreement at $f_{bh}\to 0$ is trivial, as $P_d\to 0$ in
that limit.  The agreement at larger $f_{bh}$, where $P_d\to 1$,
arises because the expected time to the first destructive encounter is
small, so evolution (and the resulting increase in concentration) is
relatively unimportant.  Nevertheless, the agreement between the
analytic formulae and simlulations is surprisingly good even for
intermediate values, $f_{bh}\sim 0.05$, where $P_d\sim 0.3-0.5$.

\subsubsection{Independence of $M_{bh}> M_{high}$}
To see how cluster evolution affects $P_d$ at higher $M_{bh}$, we
repeat the above calculations for $W_0=5$ and $M_{bh}=2M_{high}$.
Figure 4 shows the results.  They agree well with the results for
$W_0=5$ presented in the previous section.  We conclude that
evolutionary effects do not destroy the high-mass scaling derived in
the fixed cluster approximation.  This, in fact, is not surprising
since we have restricted our attention to clusters with $t_{ev}>t_H$.

\begin{figure}
\label{fig:w5_pd_m5_r16}
\epsfxsize=20pc
\epsfbox[12 138 600 726]{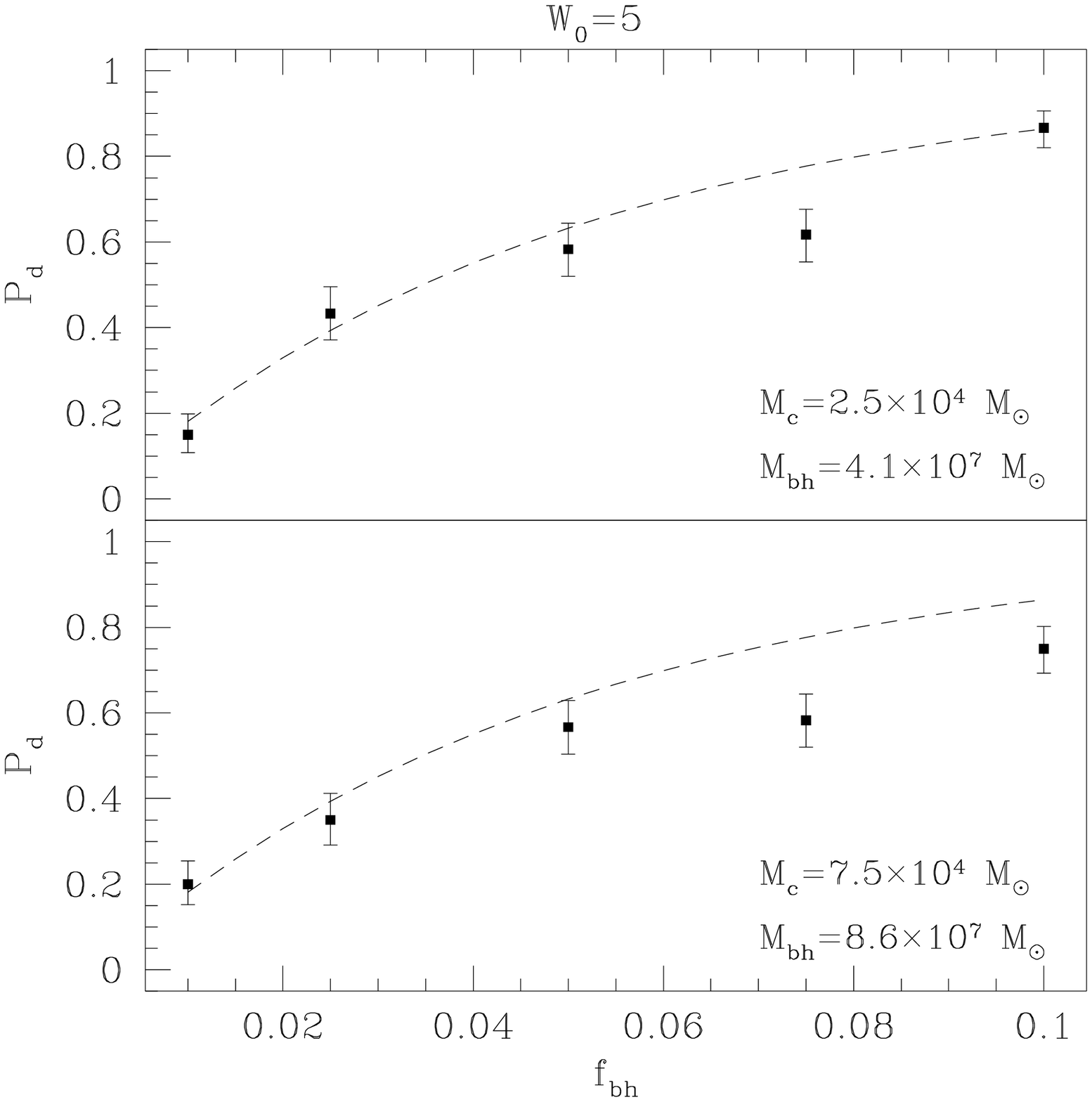}
\caption{Probabilities for 15\% encounters for $W_0=5$ clusters with
indicated masses at $16\kpc$ where $M_{bh}=2M_{high}$ for $M_{high}$
used in figures 1-3.  Solid squares with assocated error bars show the
results of Monte Carlo simulations; dashed line shows the analytic
prediction determined from equations (6) and (7).}
\end{figure}

\subsubsection{Approximate behavior for $M_{bh}<M_{high}$}
In the above analysis, we have conservatively adopted a very large
choice for $M_{high}$ to ensure that our calculations obey the
high-mass formalism in the strictest sense.  However, Arras and
Wasserman (1998) have shown that the tidal limit formula for mass loss
provides a good approximation even when $b_d<r_t$ since clusters have
fairly extended, loosely bound halos.  Typically we expect the tidal
formula to remain fairly accurate even for impacts just outside the
core radius ($b_d/r_t\sim 0.1$ for $W_0=5$).  This approximation is
worst at low concentration because the mass distribution is more
extended.  The approximation improves as the cluster evolves and
becomes more concentrated.  For example, $b_d=0.25 r_t$ encloses
roughly 50\% of the mass in the $W_0=3$ cluster and roughly 80\% of
the mass in the $W_0=7$ cluster.

For fixed $N_d$ and $V_{rel,max}=2.5 \langle V_{rel}\rangle$,
$M_{bh}\propto b_d^2$.  Thus the approximation significantly improves
the limiting $M_{bh}$ which our calculations explore.  In particular,
if we adopt $b_d=0.25 r_t$, then our calculations are valid for
$M_{bh}=0.0625 M_{high}$.  In other words the high-mass scaling
obtains for $M_{bh}<< M_{high}$ which we have defined above.  Figure 5
shows the collision probabilities calculated using this approximation.
The results agree well with the analytic predictions and indicate that
the high-mass or destructive regime obtains down to $M_{bh}\sim 10^6
M_{\odot}$.

\begin{figure}
\label{fig:w5_pd_m0.15625_r16.fig}
\epsfxsize=20pc
\epsfbox[12 138 600 726]{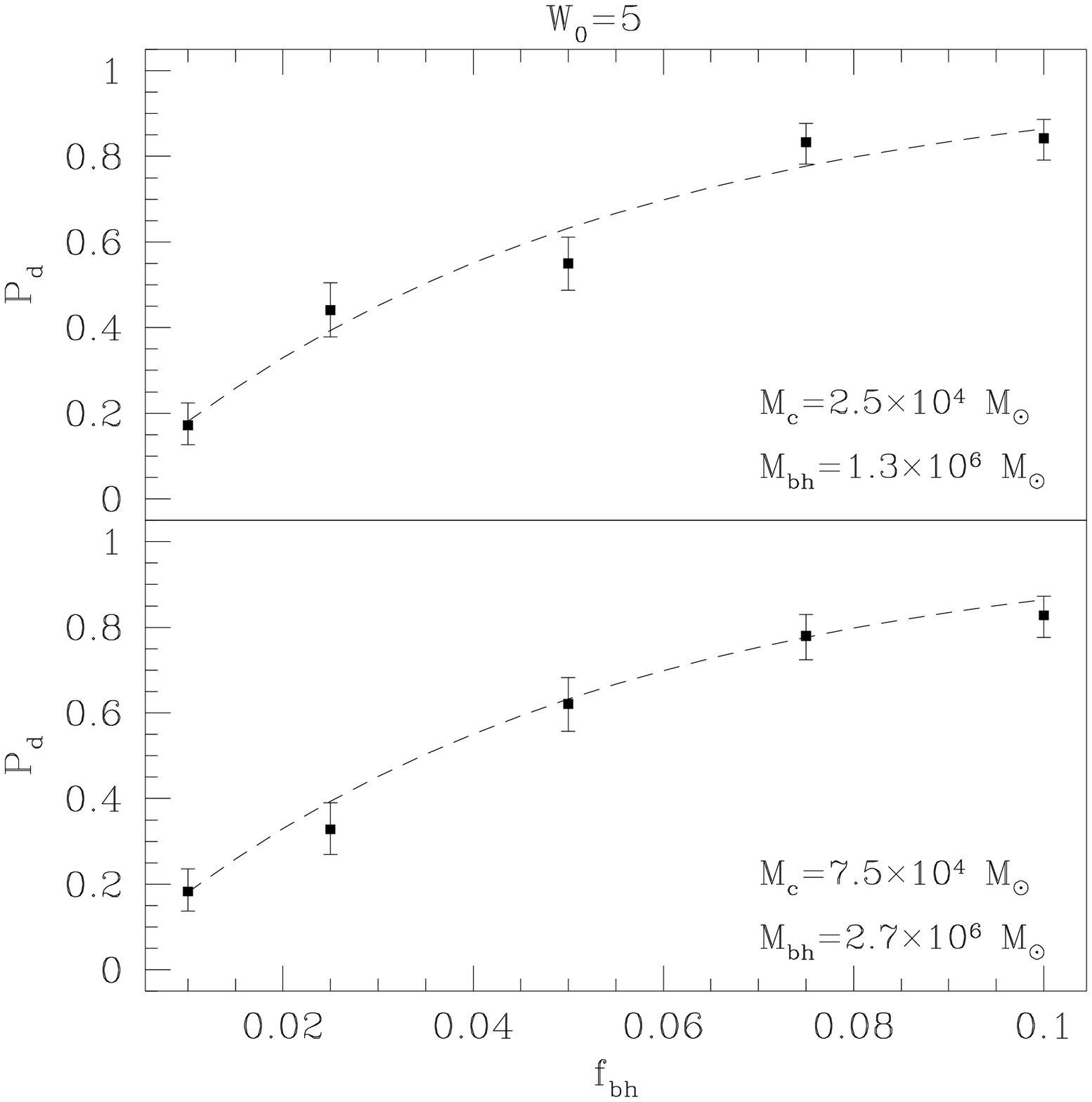}
\caption{Probabilities for 15\% encounters for $W_0=5$ clusters with
indicated masses at $16\kpc$ where $M_{bh}=0.0625M_{high}$ for
$M_{high}$ used in figures 1-3.  Solid squares with assocated error
bars show the results of Monte Carlo simulations; dashed line shows
the analytic prediction determined from equations (6) and (7).}
\end{figure}

\subsubsection{Radial scaling of collision probabilities}
The collision probabilities enumerated for $R=16\kpc$ can be scaled
approximately to larger radius by keeping $N_d$ fixed.  Since
$\rho_{halo}\propto R^{-2}$ and $K\propto R^2$ from equation
(\ref{eq:Kdef}), then $N_d$ is constant with radius for $f_{bh}\propto
R$.  The scaling is approximate because the intrinsic evolutionary
rate of a cluster varies with radius due to the variation in dynamical
time: $t_{dyn}\propto R$ for a tidally limited cluster in an
isothermal halo.  However, the above calculations fall into the regime
$t_{evap}> t_H > \Gamma_d^{-1}$, so the scaling is strong.  Additional
calculations verify the scaling.

\subsubsection{Eccentricity dependence}
Clusters on eccentric orbits in the Galaxy experience time variation
in $\Gamma_d$, the rate of destructive encounters, since the number
density of black holes varies along their orbits. The galactic tidal
force also varies along a cluster orbit; this can be accounted for
roughly by assuming that the cluster is tidally limited at the
pericenter of its orbit. The integrated number of destructive
encounters can then be written $N_d=N_d(\epsilon=0){\cal
C}(\epsilon)$, where $\epsilon=1-J/J_{max}(E)$ is a measure of the
eccentricity of an orbit with energy $E$ and angular momentum $J$, and
$N_d(\epsilon=0)$ is the number of destructive encounters for a
circular orbit with energy $E$ (and angular momentum $J_{\rm
max}(E)$).

Whether or not $N_d$ increases or decreases with increasing
eccentricity depends on a competition between a higher number of
encounters at smaller $R$, and more time spent at large $R$ near
apocenter. By numerical evaluation of ${\cal C}$, we have found that
${\cal C} \lta 1$, so that the penalty of spending more time at
apocenter with the accompanied small encounter rate is the dominant
factor. A convenient analytical formula for ${\cal C}(\epsilon)$ 
can by found by expanding the integral $N_d(t)=\int_0^t dt' \Gamma_d(t')$
for small $\epsilon$ and integrating $t'$ over an integral number of
orbital periods -- a good approximation if many orbits have
been traversed. Under these conditions we find
\begin{equation}
{\cal C}(\epsilon) \simeq 1 - \frac{3}{2} \epsilon^{1/2} + \frac{33}{24}
\epsilon  + {\cal O}(\epsilon^{3/2}).
\end{equation}

The probability of survival is larger for clusters on eccentric orbits
than for clusters on circular orbits, given $E$. For example, consider
two clusters with the same orbital energy, one on a circular orbit at
$R=16$ kpc and a second on an eccentric orbit with pericenter at
$R_p=8$ kpc; in this case $N_d$ is smaller for the eccentric orbit by
about a factor of two.  In this case, the cluster on the eccentric
orbit has a higher rate of internal evolution because of its smaller
pericenter, an effect which must also be taken into account; we
discuss qualitatively in \S 4.  Finally, we note that the average
value of the correction factor for an isotropic distribution of
angular momenta is $\langle {\cal C}(\epsilon) \rangle \simeq 0.5$ for
a fairly large range of orbital energies.

\subsection{Properties of evolved clusters}
We examine the basic properties of the clusters which do not undergo
15\% collisions.  Thus we describe the effect of the weaker encounters
on cluster evolution.  Ideally we would like to determine the Green's
function or probability amplitude for evolution from a given initial
state to a given final state (e.g. Arras \& Wasserman 1998).  Of
course, with 60 realizations per run and only the fraction
$1-P_d(f_{bh})$ not suffering strong collisions, we can only
understand the distribution of final states very approximately.  To do
so, we simply examine the mean mass and concentration of the
`surviving' clusters.

Figures 6-8 show the mean mass and concentration for clusters in the
runs discussed above.  In each case, remaining mass decreases
montonically with $f_{bh}$.  As mentioned above, the black hole flux
leads to both weak and strong encounters: clusters which do not suffer
strong collisions still lose mass through weak encounters, so the
final mass will be less than that of an isolated cluster.

The evolution of concentration $c=\log r_t/r_c$ behaves somewhat
differently in each case.  At low mass, the relaxation rate is
enchanced by weak encounters for all initial concentrations.  However,
the $W_0=3$ clusters have not yet entered core collapse so $c$
increases montonically with $f_{bh}$.  For $W_0=5$ and $W_0=7$, $c$
decreases with $f_{bh}$ because the core reaches the post-collapse
stage of expansion more rapidly due to the external heating.  

At high mass, the effect differs because of the longer intrinsic
relaxation time.  For initial $W_0=3$, $c$ decreases with $f_{bh}$
because heating has little effect on core evolution, tending only to
decrease $r_t$.  For $W_0=5$, $c$ increases with $f_{bh}$ due to the
acceleration of core evolution by external heating.  For $W_0=7$,
external heating accelerates core evolution past collapse and into
expansion, so that $c$ decreases with $f_{bh}$.

\begin{figure}
\label{fig:w3_survive_stats}
\epsfxsize=20pc
\epsfbox[12 138 600 726]{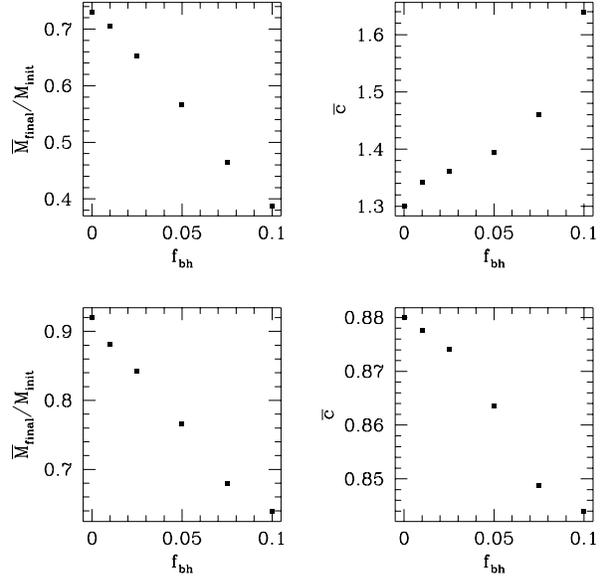}
\caption{Mean final mass and concentrations of $W_0=3$ clusters which
{\it do not} suffer 15\% collisions.  Top row shows results for
clusters with $M_{init}=2.5\times 10^4 M_{\odot}$; bottom row shows
results for clusters with $M_{init}=7.5\times 10^4 M_{\odot}$. }
\end{figure}

\begin{figure}
\label{fig:w5_survive_stats}
\epsfxsize=20pc
\epsfbox[12 138 600 726]{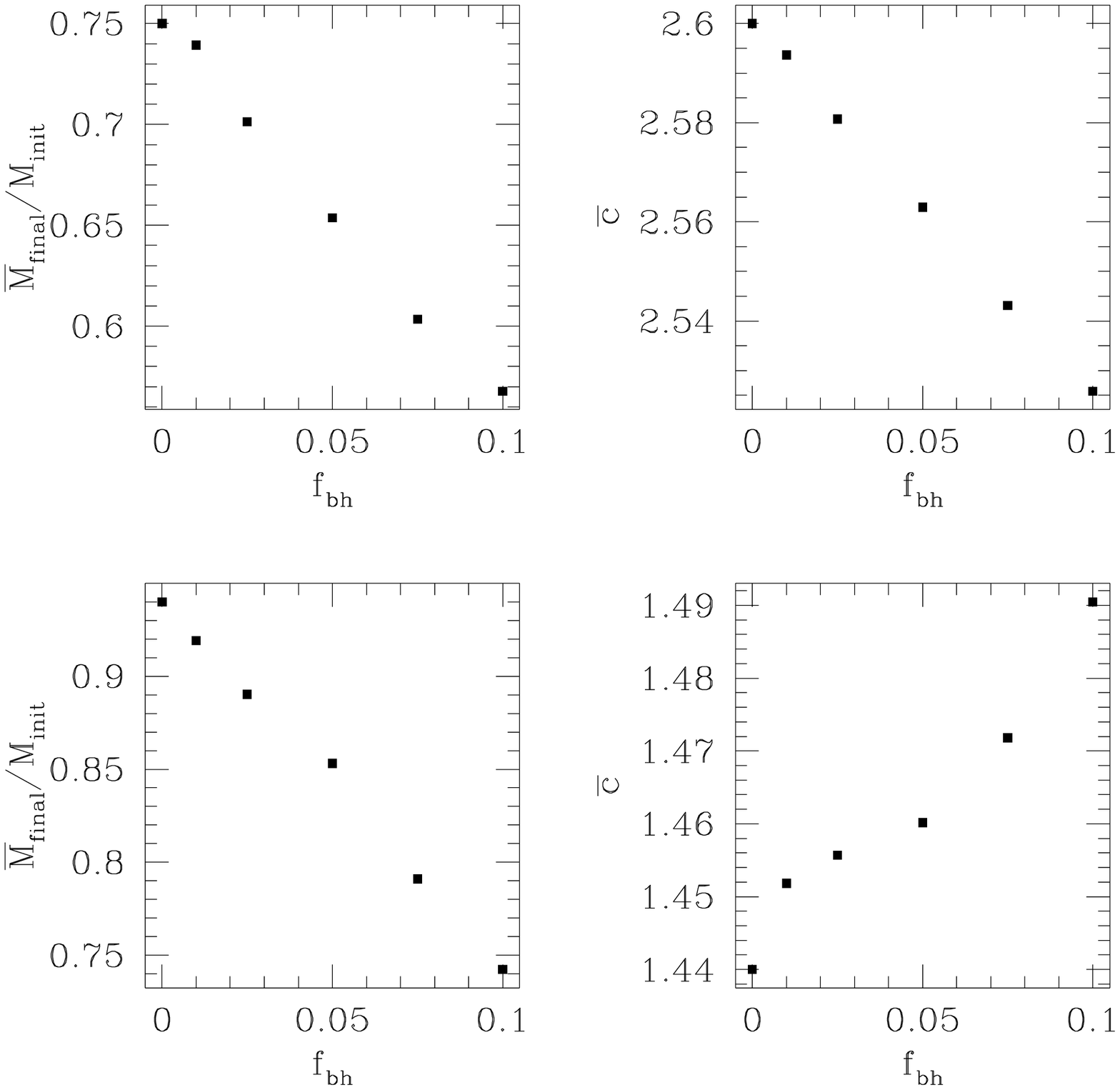}
\caption{Mean final mass and concentrations of $W_0=5$ clusters which
{\it do not} suffer 15\% collisions. Top row shows results for
clusters with $M_{init}=2.5\times 10^4 M_{\odot}$; bottom row shows
results for clusters with $M_{init}=7.5\times 10^4 M_{\odot}$.}
\end{figure}

\begin{figure}
\label{fig:w7_survive_stats}
\epsfxsize=20pc
\epsfbox[12 138 600 726]{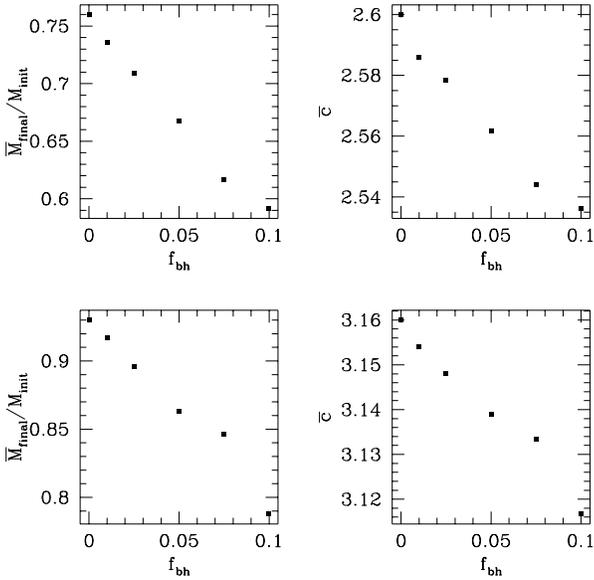}
\caption{Mean final mass and concentrations of $W_0=7$ clusters which
{\it do not} suffer 15\% collisions. Top row shows results for
clusters with $M_{init}=2.5\times 10^4 M_{\odot}$; bottom row shows
results for clusters with $M_{init}=7.5\times 10^4 M_{\odot}$.}
\end{figure}

\section{Discussion}
\label{sec:discussion}
We have examined in detail how typical globular clusters evolve in a
halo which contains a population of massive black holes with
$M_{bh}\sim 10^6 M_{\odot}$.  Our main goal was to establish
probabilities for strong collisions between clusters and black holes
in the high-mass limit using Fokker-Planck calculations in order to
combine effects of internal relaxation, binary heating and black hole
shocking.

Our results show that evolution does not radically alter collision
probabilities determined in the fixed cluster approximation (equation
\ref{eq:Pd_an}).  This result is not surprising given our approach: we
first determine initial conditions which do not lead to significant
evolution in the absence of black holes; we then include black holes
of some mass $M_{bh}$ and abundance $f_{bh}$ in calculations with
these initial conditions.  Evolution can help reduce $P_d$, but weaker
encounters tend to accelerate core collapse and evaporation by
removing mass from the halo.  The differences that do arise are
relatively small and therefore will not affect any conclusions we draw
below.

In calculating $P_d$, we have only considered clusters on circular
orbits. As discussed in \S 3.2.4, clusters on eccentric orbits have
lower collision probabilities. However, decreasing the pericenter also
shortens the evaporation time for a cluster in isolation.  For
example, Table 2 shows that a cluster with mass $M_c=2.5\times
10^4M_\odot$ has an evaporation timescale of about 20 Gyr assuming a
circular orbit of radius $R=16$ kpc; if its pericenter were $R_p=8$
kpc, its evaporation time would be about 10 Gyr. Moreover, disk
shocking becomes effective for pericenter radii $R_p\lta 8$ kpc,
further reducing the odds of survival for $M_c=2.5\times 10^4M_\odot$
even if $f_{bh}=0$. To be conservative, we could avoid these
complications by focusing on clusters of larger mass, resulting in
correspondingly larger values of the minimum black hole mass on which
we can place constraints in the high-mass limit: for a fixed
evaporation time, and assuming tidally limited clusters, we should
scale $M_c\propto R_p^{-1}$ and $M_{\rm high}\propto
M_c^{2/3}R_p^{1/3} \propto R_p^{-1/3}$, approximately.  If we choose
to be bolder, we could keep $M_c$ fixed, restrict ourselves to
acceptable values of $R_p$ (e.g. $R_p\gta 8$ kpc for $M_c=2.5\times
10^4M_\odot$) and rescale $M_{\rm high}\propto R_p^{1/3}$. Either way,
the limits on $f_{bh}$ are unlikely to change by more than a factor of
two (see \S 3.2.4).  Because we have only rigorously considered
clusters on circular orbits, we will address any remaining ambiguities
in defining $M_{high}$ in our next paper (Arras et al 1999), where we
shall consider the evolution of a realistic population of globular
clusters and work our way up from the regime of low black hole masses,
thereby obtaining limiting values of $f_{bh}$ as a function of
$M_{bh}$; where that curve asymptotes to the $M_{bh}$-independent
bound found here will determine $M_{high}$.

We have calculated $P_d$ assuming $f_d=0.15$; this restriction was
imposed by the limitations of our linear perturbation procedure. We
can scale the results to higher $f_d$ to consider collisions more
likely to disrupt the cluster, i.e., $f_d\sim 0.5$.  Here nonlinear
effects become important but the scaling approximately holds.  In
linear theory, $N_d\propto f_{bh}/\sqrt{f_d}$ from equation (6) while
$M_{high}\propto \sqrt{f_d}$.  Therefore, for fixed $f_{bh}$,
$N_d\propto 1/\sqrt{f_d}$ and the values of $P_d$ calculated above
decrease correspondingly for $f_d>0.15$.  For fixed $N_d$ and $P_d$,
$f_{bh}\propto \sqrt{f_{bh}}$. (Note that if $N(f_d^\prime)\sim 1$,
then for $f_d<f_d^\prime$, $N_d(f_d)\sim\sqrt{f_d^\prime/f_d)}$, which
would only result in a fractional mass loss $\sim\sqrt{f_d^\prime
f_d}<f_d^\prime$; large fractional mass loss in a single encounter is
always more likely than in numerous encounters each with smaller
fractional mass loss.)

We can also scale to different galactocentric radii using the
approximate relationships $f_{bh}\propto R$ (see \S 3.2.3) and $M_{\rm
high}\propto f_d^{1/2}M_c^{2/3}R^{1/3}$, which becomes $M_{\rm
high}\propto f_d^{1/2}R^{-1/3}$ for fixed evaporation time (see \S
3.2.4); including disk shocking -- which we have neglected here --
ought to raise the value of $M_{\rm high}$ somewhat because only
higher mass clusters survive. The constraints on $f_{bh}$ become
stronger for smaller galactocentric radii, but we expect $f_{bh}$ to
be nonuniform as a consequence of dynamical friction, so limits at
relatively large $R$ are easiest to interpret.

\subsection{Interpreting the collision probabilities}
Our calculations in smooth halos, $f_{bh}=0$, combined with the
observational picture serve as a guide for interpreting the collision
probabilities.  Observationally, the similarity of globular cluster
luminosity functions over a range of galaxy environments (e.g. Harris
1991) may reflect the formation process and suggests that cluster
populations are relatively unevolved for $M_c\sim 10^5 M_{\odot}$.
Moreover, the flatness of the distribution of globular cluster
luminosities, $dN/dL_c$ -- and hence $dN/dM_c$ -- at low $L_c$ in the
Milky Way (e.g. Ashman \& Zepf 1998) suggests that globular clusters
with $M\lta 10^5\msun$ are not whittled away rapidly, but are
relatively long-lived since, otherwise, the rate of destruction would
far exceed the rate of production and the distribution would fall off
drastically.  Our calculations with $f_{bh}=0$ appear to be reasonably
consistent with this expectation.  In particular, these calculations
show that clusters on circular orbits at $16\kpc$ survive beyond a
Hubble time for $M\gta 10^4 M_{\odot}$, incurring roughly 25\% mass
loss for $M_c=2.5\times 10^4 M_{\odot}$ and roughly 5\% mass loss for
$M_c=7.5\times 10^4 M_{\odot}$.  A stringent view, therefore, requires
that halo black holes leave these clusters relatively unscathed.

Our numerical simulations show that the probability of a collision
with $f_d=0.15$ is about 30-40\% for $f_{bh}\approx 0.025$, about 50\%
for $f_{bh}\approx 0.05$, and about 80\% for $f_{bh}\approx 0.1$;
assuming that the collision probability in this range of $f_{bh}$ is
excessive, we can rule out $M_{bh}\gta 1.3\times 10^6M_\odot$ for
$M_c=2.5\times 10^4M_\odot$ and $M_{bh}\gta 2.7\times 10^6M_\odot$ for
$M_c=7.5\times 10^4M_\odot$ for $f_{bh}\gta 0.1$.  From the scaling
$N_d\propto f_{bh}/\sqrt{f_d}$, equal collision probabilities for
$f_d=0.5$ imply the values $f_{bh}=0.05$, $f_{bh}=0.09$ and
$f_{bh}=0.18$, respectively; since $M_{high}\propto\sqrt{f_d}$ (see
Section 2.2, especially eq. [3]), these limits apply to $M_{bh}\gta
2.4\times 10^6\msun$ for $M_c=2.5\times 10^4\msun$ and $M_{bh}\gta
4.9\times 10^6\msun$ for $M_c=7.5\times 10^4\msun$.  These are
somewhat higher but still very restrictive.  

Black hole collisions do not only destroy clusters, but also whittle
away their masses. For low $\Fd\Nd$, the fractional mass loss endured
by surviving clusters is approximately $\Fd\Nd$ in the tidal limit
(Figs. 6-8 and Arras \& Wasserman 1999).  Thus, we expect that for
large $\Fd\Nd$, the mean mass per cluster declines like
$\exp(-\Fd\Nd)$, and the total mass in clusters declines like
$\exp[-(1+\Fd)\Nd]$ when cluster destruction is taken into account, in
the tidal limit. To a first approximation, the evolution of the
distribution of clusters must account for a steady advection downward
in mass as well as destruction. If $N(M,t)dM$ is the number of
clusters with masses between $M$ and $M+dM$ at time $t$, then 
\begin{equation}
{\partial N(M,t)\over\partial t}=-\gd N(M,t)
+\gd\Fd{\partial[N(M,t)M]\over\partial M}
\label{eq:advectidal}
\end{equation}
represents a simple advection-destruction model appropriate for the
tidal limit. The solution to this equation is
\begin{equation}
N(M,t)=\exp[-\gd(1-\Fd) t]N(M\exp(\Fd\gd t),0).
\label{eq:soltidal}
\end{equation}
According to this solution, the total number of clusters declines
$\propto\exp(-\gd t)=\exp(-\Nd)$ and the total mass remaining in
clusters after time $t$ is
$\propto\exp[-\gd(1+\Fd)t]=\exp[-(1+\Fd)\Nd]$.

For a given black hole mass $\Mbh$, the tidal approximation holds up
to some maximum cluster mass, $\Mt$; if attention is restricted to the
tidal approximation, which considers only collisions with impact
parameters outside $r_t$, this maximum mass is $\Mt\sim({\rm
a\,few})\times 10^4\msun$ for $\Mbh=10^6\msun$.

Black hole collisions still destroy and whittle away clusters with
masses above $\Mt$. Consequently, clusters destroyed and chiselled
away at masses below $\Mt$ are replaced, to a degree, by clusters
originally at masses above $\Mt$.  Our calculations in the tidal limit
cannot describe this evolutionary process; to do so requires
calculations of what happens as a consequence of black hole collisions
at $b<r_t$, which we have excluded in this paper (but are in the midst
of calculating, and will publish separately). However, from earlier
work (e.g. Arras \& Wasserman 1999), we already know that the tidal
approximation continues to describe the mean mass loss by a cluster to
within 20-30\% for somewhat smaller impact parameters,
$b\gta(0.1-0.2)r_t$.  A quantitatively reasonable interpolation
formula for the fractional mass loss of a cluster due to a collision
with a black hole passing within impact parameter $b$ of the center of
a cluster at relative speed $\vrel$ is 
\begin{equation}
f(b,\vrel)={f(0,V_c)(V_c/\vrel)^2\over [1+(b/b_0)^2]^2},
\label{eq:interpmodel}
\end{equation}
where $f(0,V_c)$ is the fractional mass loss at zero impact parameter
for relative velocity $\vrel=V_c$; for a tidally limited cluster at
galactocentric radius $r$, $f(0,V_c)\propto\Mbh^2/M^{4/3}r^{2/3}$.
The numerical value of $f(0,V_c)$ can be found using the results of
Arras \& Wasserman 1999 (see fig.3). The value of $b_0$ is fixed by
requiring the rate of destructive encounters found using
eq.\ref{eq:interpmodel} give the correct tidal limit.

For a given black hole mass, there is a new critical mass $\Md$ above
which clusters become considerably more immune to destruction. Typical
penetrating encounters are nondestructive when the cluster mass is
large enough that $f(0,V_c) < f_d$. Using eq. (\ref{eq:interpmodel}),
we estimate
\begin{equation}
\Md\approx 2\times 10^5\msun\biggl({\Mbh\over
10^6\msun}\biggr)^{3/2}\biggl({8\,{\rm kpc}\over R}
\biggr)^{1/2}\biggl({0.5\over f_d}\biggr)^{3/4}.
\label{eq:md}
\end{equation}
The evolution of clusters with $M>\Md$ in the face of black hole
collisions is described poorly by the tidal approximation for two
reasons. First, the rate of destructive encounters is low, because
$f(0,V_c)$ drops below $\Fd$ for $M>\Md$. Second, the rate at which
cluster advect downward in mass is slowed because of the same cutoff
in fractional mass loss.  We can generalize eq. (\ref{eq:advectidal})
to account for the different behavior at masses above and below $\Md$;
the resulting equation is 
\begin{eqnarray}
{\partial N(M,t)\over\partial t} & = & -\gd I(M/\Md)N(M,t) 
\nonumber \\ & + & 
 \gd\Fd{\partial[N(M,t)MH(M/\Md)]\over\partial M},
\label{eq:advectinterp}   
\end{eqnarray}
where, from the scalings implied by eq. (\ref{eq:interpmodel}), we
infer that $I(z)\to 1$ for $z\ll 1$ and $I(z)\to 0$ for $z\gg 1$, and
$H(z)\to 1$ for $z\ll 1$ and $H(z)\sim z^{-2/3}$ for $z\gg 1$.  (The
exact $M/\Md$ dependences can be computed from
eq. [\ref{eq:interpmodel}] given the distribution of $V$.)

The tidal limit is recovered for essentially the entire range of
globular cluster masses for sufficiently large $\Mbh$. The most
massive Galactic globular cluster, $\omega$ Cen, has $M_c\approx
2.4\times 10^6\msun$ (e.g. Ashman \& Zepf 1998, Mandushev et
al. 1991), and $\Md>2.4\times 10^6\msun$ for $\Mbh>5.2\times
10^6\msun(R/8\,{\rm kpc})^{1/3} (0.5/\Fd)^{1/2}$ from
eq. (\ref{eq:md}).  Since cluster masses decrease with time as a
result of black hole perturbations, $\Mbh$ must be somewhat larger
still for the tidal approximation to hold for all time.

According to the advection-destruction solution,
eq. (\ref{eq:soltidal}), the cluster mass distribution evolves
self-similarly with time in the tidal limit, which means that the
initial conditions must have resembled the distribution seen today,
except shifted to larger mass.  Thus, we cannot use the shape of the
distribution to derive a bound on $f_{bh}$, without additional
information on the initial mass function for clusters, and without
including other processes, such as evaporation and (at $R\lta 8$ kpc)
disk shocking.
\footnote{Since the evaporation time scales $\propto M$, the rate of
evaporative mass loss, $\dot M_{ev}$, is approximately independent of
mass, and the tidal limit solution with evaporation is
$N(M,t)=\exp[-(1-\Fd)t]N(M_i(t),0)$, with $M_i(t)=M\exp(\Fd\gd
t)+(\dot M_{ev}/\gd\Fd) [\exp(\Fd\gd t)-1]$. Evaporation can be
included similarly in solving eq. (\ref{eq:advectinterp}). In deriving
limits on $\fbh$, we neglect evaporation, which would make our
constraints slightly tighter.}  Requiring that the total mass lost by
clusters not exceed about 100 times the total mass presently contained
in them, so the halo star population is not primarily due to ejecta
from clusters (e.g. Klessen \& Burkert 1996 and references therein),
implies $\Nd\lta 3.1$, or $\fbh\lta 0.36(R/16\,\kpc)$, for
$\Fd=0.5$. This limit is fairly conservative, since metallicity
differences suggest that the bulk of halo field stars did not
originate from globular clusters (Harris 1991). Moreover, clusters on
nearly circular orbits are somewhat likelier to be disrupted than
clusters on elongated orbits, so we should expect ejected stars to
have a tangentially-biased velocity ellipsoid in the inner Galaxy,
which is not observed for the halo stars (Beers \& Sommer-Larson
1995).

For smaller $\Mbh$, for which $\Md$ falls in the range of present (and
past) cluster masses, we can use eq. (\ref{eq:advectinterp}) to get a
rough idea of how the cluster distribution function evolves as a
consequence of bombardment by massive black holes.  (We shall present
more realistic and accurate models, as well as statistical analyses,
in a subsequent paper.) Qualitatively, clusters with masses below
$\Md$ ought to lose mass and be destroyed rapidly.  Clusters with
masses above $\Md$ are less prone to destruction, and lose mass more
slowly. Although the advection of clusters from masses above $\Md$ to
below $\Md$ ought to replenish the supply of low mass clusters lost to
destructive collisions, the slowness of the process results in an
overall truncation of $N(M,t)$ at low values of $M$.

To explore the stability of the presently-observed mass distribution
of clusters, we used eq.  (\ref{eq:advectinterp}) to compute the
evolution of a distribution that is originally 
\begin{equation}
N(M,0)={1\over\Mz(1+M/\Mz)^2},
\label{eq:nm0}
\end{equation}
with $\Mz=3\times 10^5\msun$, up to a maximum mass $M_{\rm
max}=10^7\msun$; the calculations assumed $\Fd=0.5$ and $\Md=6\times
10^5\msun$ (corresponding to $\Mbh=2\times 10^6\msun$).  The original
cluster mass distribution given by eq. (\ref{eq:nm0}) is flat below
$\Mz$ and $\propto M^{-2}$ above, in reasonable agreement with the
observed luminosity function for globular clusters in the Milky Way
(and other galaxies), assuming a constant $M/L\approx 3$ (e.g. Ashman
\& Zepf 1998). The results for $MN(M,t)$, the distribution of clusters
in $\ln M$, is shown in Fig. 9a; the logarithmic slope $\partial\ln
N(M,t)/\partial\ln M$ is shown in Fig. 9b. These figures illustrate
the truncation of the distribution at low masses, and show how the
shape of the distribution tends to evolve with time. From Fig. 9,b, we
see that the shape begins to deviate significantly from the original
one after $\gd t=\Nd\approx 2$ or $3$. Although this result is
preliminary, we expect that more sophisticated analysis will lead to a
similar bound.

We can also examine the evolution of the globular cluster system from
presumed initial conditions using eq. (\ref{eq:advectinterp}). As in
the tidal case, we remark that the evolutionary sequences computed
here only include the effects of black hole perturbations, not the
better established effects known to operate, such as evaporation and
disk shocking. For illustrative purposes, we adopted $N(M,0)\propto
M^{-2}$ for $10^3\msun\leq M\leq 10^7\msun$ (e.g.  McLaughlin
1999). The results are shown in Fig.  9c,d. As can be seen, $N(M,t)$
evolves to a form reminiscent of today's cluster mass distribution
over a time span $\gd t\sim 10$, which corresponds to $\approx 10$ Gyr
for $\fbh\approx 1.1(R/16\,\kpc)$. However, it is also clear from this
figure, and from Fig. 9a,b, that the shape of $N(M,t)$ evolves
somewhat more rapidly, on a timespan $\gd\Delta t\sim 2$ or $3$, so
the observed distribution is not long-lived. We note that the total
mass in clusters drops by about a factor of ten between $\gd t=0$ and
$\gd t=10$ for this model, which, although substantial, does not
violate any observational constraints (cf. Klessen \& Burkert 1996).

\begin{figure}
\label{fig:evolution}
\epsfxsize=20pc
\epsfbox[12 138 600 726]{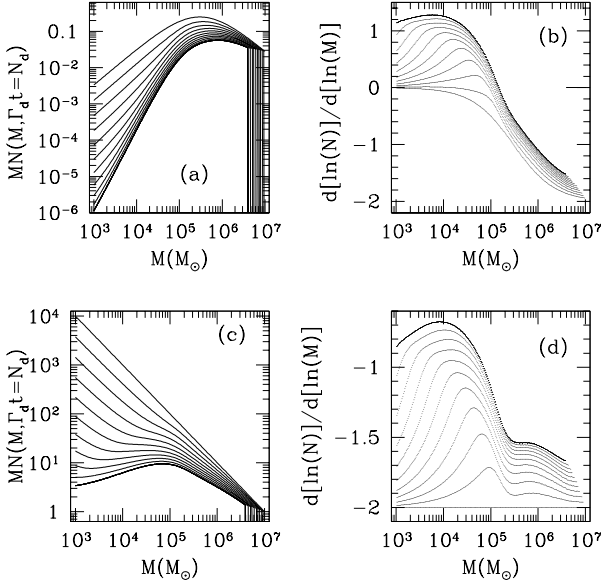}
\caption{Panels (a-d) show the cluster population mass function and
its power law index for a range of times.  The initial condition
$N(m,0)=(1+M/M_0)^{-2}/M_0$ was used in figures (a) and (b), and the
initial condition $N(m,0)=1/M^2$ was used in figures (c) and (d). The
eleven curves in each figure correspond to the times $\Gamma_d
t=N_d=0,1,...,10$. In figures (a) and (c), time increases as the
curves go from top to bottom, while in figures (b) and (d) the
opposite is true. The sharp edges seen at large masses occur where the
mass function is zero in the advection/destruction model.}
\end{figure}

From these preliminary investigations, we conclude that while it is
possible to find initial conditions $N(M,0)$ that evolve to the
observed distribution of cluster masses in $\sim 10$ Gyr even for
$\fbh\sim 1$, the evolutionary timescale is rather short, so the
distribution is rather unstable. Although a more precise treatment of
the bounds on $\fbh$ that can be derived from requiring stability of
the cluster mass distribution needs to be done (and will be presented
by us elsewhere), we are confident that the observed distribution is
unstable unless $\gd t=N_d\lta 3$ everywhere in the halo. Adopting
this rather conservative limit, we conclude provisionally that
stability of the cluster mass function requires $\fbh\lta
0.34(R/16\,{\rm kpc})$.

We therefore propose two different bounds on $\fbh$. The less
restrictive of the two is the stability bound derived above. For
direct comparison with the disk heating bound, we shall rescale all
constraints to $R=8\,\kpc$; then stability implies $\fbh\lta
0.17(R/8\,{\rm kpc})$. This bound applies, strictly speaking, to
$\Mbh=2\times 10^6\msun$. At much larger black hole masses, where the
tidal approximation is valid, we found a similar bound based on mass
loss from clusters to the halo, and we adopt $\fbh\lta 0.17(R/8\,{\rm
kpc})$ for all $\Mbh\gta 2\times 10^6\msun$.  The more restrictive
bound is obtained by assuming that a hypothetical cluster with
$M_c=2.5\times 10^4 M_{\odot}$ on a circular orbit at 16 kpc should
have no worse than a 50\% chance of survival.  Using the scalings
given above to compute bounds at $R=8\,\kpc$, we find $f_{bh}\lta
0.025$ for $f_d=0.15$ and $f_{bh}\lta 0.05$ for $f_d=0.5$.  If we keep
$M_c=2.5\times 10^4M_\odot$ fixed, then these limits apply to
$M_{bh}\gta 1\times 10^6M_\odot$ and $M_{bh}\gta 2\times 10^6M_\odot$,
respectively. Since the evaporation time falls to about 10 Gyr at
$R=8$ kpc for $M_c=2.5\times 10^4M_\odot$ and $f_{bh}=0$, we might
prefer to keep the evaporation timescale fixed at $20\Gyr$, in which
case our derived limits on $f_{bh}$ apply to slightly larger black
hole masses, $M_{bh}\gta 2\times 10^6M_\odot$ and $M_{bh}\gta 3\times
10^6 M_\odot$ for $f_d=0.15$ and $f_d=0.5$, respectively. For our
conservative bounds, we adopt $f_{bh}\lta 0.05$ for $M_{bh}\gta
3\times 10^6M_\odot$; for our most stringent bounds, we adopt
$f_{bh}\lta 0.025$ for $M_{bh}\gta 1\times 10^6M_\odot$.  Figure 10
shows these limits along with the upper limit
\begin{equation}
M_{\rm bh}<4.4\times 10^7\msun\biggl({8\over\ln
\Lambda}\biggr)
\biggl({R\over 8\,{\rm kpc}}\biggr)^2,
\end{equation}
for $\ln\Lambda=8$ imposed by requiring that dynamical friction be
incapable of dragging black holes inward in 10 Gyr (see equation
[7-27] in Binney \& Tremaine 1987), and the disk heating constraint
$f_{bh}M_{bh}\lta 2\times 10^6\msun$ (Lacey \& Ostriker 1985).  The
figure shows that the globular cluster constraint forbids considerable
portions of $M_{bh}-f_{bh}$ space allowed by disk heating.

\begin{figure}
\label{fig:constraint}
\epsfxsize=20pc
\epsfbox[12 138 600 726]{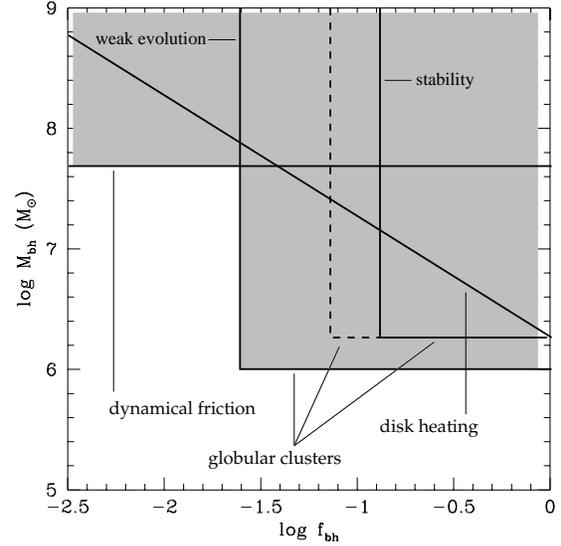}
\caption{The constraints on the mass and fraction of black holes at
$8\kpc$ in the halo.  The solid horizontal line shows the upper limit
on $\log M_{bh}$ from dynamical friction.  For $\log M_{bh}\gta 7.7$,
black holes spiral into the Galactic center.  The solid diagonal line
shows the disk heating constraint: $\log f_{bh}M_{bh}\gta 6.3$
overheats the disk.  The solid box labeled {\it stability} shows the
constraint obtained from the stability requirement for the globular
cluster population, $\log f_{bh}<-0.77$ for $\log M_{bh}\gta 6.3$.
The dashed box shows the more conservative constraint for the weak
cluster evolution hypothesis $\log f_{bh}\lta -1.3$ for $\log
M_{bh}\gta 6.3$ and the solid box labeled {\it weak evolution} shows
the most stringent bound, $\log f_{bh}<-1.6$ for $\log M_{bh}\gta 6$.
The shading indicates regions which are disfavored by the combination
of constraints.}
\end{figure}

\section{Acknowledgements}
We would like to thank Scott Tremaine and the referee, Oleg Gnedin,
for detailed discussion and many helpful comments.  This work was
supported by the Fund for Astrophysical Research and NASA NAG 5-3097.

\appendix
\section{Second-order change in DF in impulsive encounter}
We derive the second-order change in the cluster DF due to an
impulsive, tidal encounter with a perturber.  As mentioned above, the
resulting expression is equivalent to a Fokker-Planck equation for the
change in the DF and therefore consists of an advection term and a
diffusion term. Our derivation reveals an error in previous treatments
(Kundic \& Ostriker 1995; Gnedin \& Ostriker 1997; Murali \& Weinberg
1997a) which results from improper integration over velocity
coordinates in projecting the advection and diffusion coefficients to
energy space. As shown in the derivation and discussion below, the
problem arises because one cannot integrate over the entire range of
angular coordinates in velocity space: failure to restrict the
integration domain corresponds to including fictitious transitions to
bound states from unbound states that are, in reality, unoccupied
initially. Mathematically, there is a
$\theta$-function implicit in the kinetic equation which has
previously been ignored.

For any impulsive enounter with a perturber, the equation,
\begin{eqnarray}
f_{new}(\bfr',\bfv')&=&\int d\bfr d\bfv \delta(\bfr-\bfr')
        \delta(\bfv-\bfv'+\Delta\bfv(\bfr)) f(\bfr,\bfv) \nonumber\\
	&=&f(\bfr',\bfv'-\Delta\bfv(\bfr'));
\label{eq:fnew}
\end{eqnarray}
gives an exact expression for the new DF, $f_{new}$, in terms of the
old DF $f$.  The $\delta$-function in time indicates that the
perturbation is impulsive; the $\delta$-function in position indicates
that particles do not move during the perturbation; and the
$\delta$-function in velocity defines the position-dependent velocity
impulse (with respect to center of mass), $\Delta\bfv(\bfr)$, imparted
to a particle by the perturbation.  Consequently, the new DF is the
old DF with velocity bins shifted according to the position-dependent
velocity impulse.

The resulting mass loss
\begin{equation}
\delta M=\int_{|\bfv'|>v_e(r')} d\bfr' d\bfv' f_{new}(\bfr',\bfv'),
\label{eq:mass_loss}
\end{equation}
is the integral over all particles whose new velocity is greater than
the escape velocity $v_e(r)$.  Substituting equation (\ref{eq:fnew})
for $f_{new}$, we can show that 
\begin{equation}
\delta M=\int d\bfr\int d\bfv\Theta(|\bfv+\Delta\bfv(\bfr)|-v_e(r)) f(r,v),
\end{equation}
where we have dropped the primes on the spatial coordinates.  This is
precisely equation (39) of Chernoff, Kochanek \& Shapiro (1987)
(hereafter CKS); therefore the present treatment of individual
collisions is equivalent to that used by Arras \& Wasserman (1998).

To use a one-dimensional, phase-space method like the standard
Fokker-Planck calculation, we must project the new DF into energy
space by integrating over all other coordinates.  The projection
\begin{eqnarray}
16\pi^2 P(E) f_{new}(E)\equiv
\int d\bfr d{\bf\Omega}_v \sqrt{2(E-\Phi)}f_{new}(\bfr,\bfv)=\nonumber\\
\int_{\vert \bfv-\Delta\bfv\vert < v_e(r)} d\bfr d{\bf\Omega}_v
\sqrt{2(E-\Phi)}f(\bfr,\bfv-\Delta\bfv(\bfr)) 
\label{eq:fnewE}
\end{eqnarray}
where the phase-space volume
\begin{equation}
P(E)=\int dr r^2 \sqrt{2[E-\Phi(r)]}.
\end{equation}
The second integral in equation (\ref{eq:fnewE}) has a limited range
of integration because the perturbation can transport unoccupied,
unbound states to bound states: areas where the unperturbed DF
vanishes, i.e. $E>E_t$, define the excluded regions.  To reiterate,
previous treatments have overlooked this subtlety in deriving the
kinetic equation, thereby obtaining a factor of 2 overestimate in the
mass loss.

To derive the second-order change, we expand $f_{new}$ in a Taylor
series about the unperturbed DF:
\begin{eqnarray}
f_{new}(\bfr,\bfv)=f(\bfr,\bfv-\Delta\bfv(\bfr))
\approx f(\bfr,\bfv)-\pdrv{f}{\bfv}\cdot\Delta\bfv(\bfr) \nonumber\\
	+\half\Delta\bfv(\bfr)\cdot
        {\partial^2 f\over \partial\bfv\partial\bfv}\cdot\Delta\bfv(\bfr).
\label{eq:fnewexp}
\end{eqnarray}
This has the standard form of the velocity-space Fokker-Planck
equation.  Substitution into equation (\ref{eq:fnewE}) yields an
equation for the change in the DF, $\delta f$.  

We will now briefly outline the calculation of $\delta f$. Let the
direction of the relative velocity be along the $z-$axis and the position
of a star in the cluster given by spherical radius $r$ and the cosine of 
the angle with respect to the $z$ axis, $\mu=\cos\theta$. The magnitude
of the tidal velocity kick with respect to the center of mass of the 
cluster is then [see, e.g., Arras \& Wasserman (1998)]
\begin{eqnarray}
\Delta v(\bfr) & = & \frac{2GM_{bh}r\sqrt{1-\mu^2}}{b^2V_{rel}}.
\end{eqnarray}
The region of phase space for which $\vert \bfv-\Delta\bfv\vert < v_e(r)$
has been discussed in both CKS and Arras \& Wasserman (1998). They find
that, depending on the energy $E$ and the position $r$ and $\mu$, 
the velocity angle cosine $\mu_v=\bfv \cdot \Delta\bfv/ v \Delta v$ 
may only extend over a restricted interval instead of $(-1,1)$. Hence, 
spherical symmetry is broken and the first order term in $\Delta v$ in
equation \ref{eq:fnewexp} no longer integrates to zero. The end result is
that there will be two different mathematical forms for $\delta f$ 
depending on the energy $E$. Define the critical energy 
\begin{eqnarray}
E_{\rm crit} &= & E_{\rm max} - \frac{2GM_{bh}r_{\rm peak}}{b^2V_{rel}}
v_e(r_{\rm peak})
\end{eqnarray}
where $r_{\rm peak}$ is the radius at which $rv_e(r)$ reaches a maximum.
For energies $E<E_{\rm crit}$, the velocity cosine $\mu_v $ is in 
$(-1,1)$ for all values of $r$ and $\mu$ and
one would obtain the ``standard" results for $\delta f$. For 
$E>E_{\rm crit}$, on the other hand, $\mu_v$ runs over a restricted range
and a different mathematical expression for $\delta f$ is obtained.

It is convenient to write $\delta f$ in the quasilinear form
\begin{equation}
\delta f(E)  =  \left[ 16 \pi^2 P(E) \right]^{-1} \frac{ d F(E)}{dE}.
\end{equation}
For energies $E<E_{\rm crit}$, we find
\begin{eqnarray} 
F(E) & = &  \frac{64\pi^2}{9}  \frac{df(E)}{dE}
\left( \frac{GM_{bh}}{b^2V_{\rm rel}} \right)^2
\nonumber \\ & \times & 
\int_0^{\phi^{-1}(E)} dr r^4 \left( 2[E-\phi(r)] \right)^{3/2},
\end{eqnarray}  
where $\phi^{-1}(E)$ is the apocenter of a radial orbit with energy $E$.
For $E>E_{\rm crit}$, a more complicated expression results, given by
\begin{eqnarray}
&& F(E)  =
+ \frac{64\pi^2}{9} \frac{df(E_{max})}{dE}
\left( \frac{GM_{bh}}{b^2V_{rel}} \right)^2
\int_0^{r_t} dr r^4 v^3_{e}(r)
\nonumber \\ &&
- 8\pi^2 \frac{ df(E_{max})}{dE}
\int_{r_{min}(E)}^{r_{max}(E)} dr r^2 v_{e}(r)
\nonumber \\ && \times
\left[ + \frac{1}{2} (E_{max}-E)^2 \mu_0(r,E)
\right. \nonumber \\ && \left.
- \frac{1}{4} v_{e} \frac{2GM_{bh}r}{b^2V_{rel}}(E_{max}-E)
\left\{ \frac{\pi}{2} - \theta_0(r,E) + \frac{1}{2}\sin(2\theta_0(r,E)
\right\}
\right. \nonumber \\ && \left.
- \frac{1}{6}\frac{1}{v_{e}(r)} \frac{b^2V_{rel}}{2GM_{bh}r}
(E_{max}-E)^3\left(\frac{\pi}{2} - \theta_0(r,E) \right)
\right. \nonumber \\ && \left.
+ \frac{1}{6}v_{e}^2(r) \left(\frac{2GM_{bh}r}{b^2V_{rel}} \right)^2
\left\{ \mu_0(r,E) - \frac{1}{3} \mu_0^3(r,E) \right\} \right].
\end{eqnarray}
Here $E=v^2/2+\phi(r)$ is the energy (per unit mass), 
$E_{max}=\phi(r_t)$ and 
$E_{min}=\phi(0)$, $r_t$ is the tidal radius of the cluster, $f$ is the
pre-collision distibution function, $r_{max}$ and $r_{min}$ are defined by
the equation
\begin{equation}
\frac{b^2V_{rel}}{2GM_{bh}}
\left( E_{max}-E \right) = v_{e}(r_{min,max}) r_{min,max},
\end{equation}
and the cosine $\mu_0(r,E)=\cos(\theta_0(r,E))$ is defined by
the equation
\begin{equation}
\left( 1 - \mu_0^2(r,E) \right)^{1/2}= 
 \frac{b^2V_{rel}}{2GM_{bh}r} \frac{1}
{v_{e}(r)} (E_{max}-E).
\end{equation}
For most of the range of $E$, the above equation gives the same result as
if there were no restrictions on phase space. Only for energies within
roughly $\Delta \bfv/v$ of $E_{max}$ do the restrictions make a 
difference. However, the flux at the boundary, which is the change in the
mass of the cluster is significantly altered; we find
\begin{equation}
F(E_{max}) = \Delta M = \frac{32\pi^2}{9} \frac{df(E_{max})}{dE}
\left( \frac{GM_{bh}}{b^2V_{rel}} \right)^2
\int_0^{r_t} dr r^4 v^3_{e}(r).
\end{equation}
This is exactly a factor of two smaller than the answer obtained from 
ignoring the phase space restrictions.

For reference, we give the value of $K$ used in equation (2):
\begin{equation}
K = \frac{32\pi^2}{9} G^2 \frac{1}{M} 
\left| \frac{df(E_{max})}{dE} \right|
\int_0^{r_t} dr r^4 v^3_{e}(r).
\end{equation}

The above procedure is inherently linear and, therefore, has only a
limited range of validity.  The method fails because the new DF,
$f_0+\Delta f$, becomes negative as we increase the mass loss.  We
find that negative excursions in the new DF become unacceptably large
for $dM/M\gta 0.15$.

\end{document}